# TEM at Millikelvin Temperatures: Observing and Utilizing Superconducting Qubits

Hiroshi Okamoto[a,*], Reza Firouzmandi[b], Ryosuke Miyamura[a], Vahid Sazgari[b], Shun Okumura[a], Shota Uchita[a], and Ismet I. Kaya[b,c]

[a] Department of Intelligent Mechatronics, Akita Prefectural University, 84-4, Aza Ebinokuchi, Tsuchiya, Yurihonjo, Akita 015-0055, Japan

[b] Faculty of Engineering and Natural Sciences, Sabancı University, Orhanli-Tuzla, Istanbul 34956, Turkey

[c] SUNUM, Nanotechnology Research and Application Center, Sabanci University, Istanbul, 34956, Turkey

*Corresponding author.
*E-mail address:* okamoto@akita-pu.ac.jp

ABSTRACT

We present a case for developing a millikelvin-temperature transmission electron microscope (TEM). We start by reviewing known reasons for such development, then present new possibilities that have been opened up by recent progress in superconducting quantum circuitry, and finally report on our ongoing experimental effort. Specifically, we first review possibilities to observe a quantum mechanically superposed electromagnetic field around a superconducting qubit. This is followed by a new idea on TEM observation of microwave photons in an unusual quantum state in a resonator. We then proceed to review potential applications of these phenomena, which include low dose electron microscopy beyond the standard quantum limit. Finally, anticipated engineering challenges, as well as the authors' current ongoing experimental effort towards building a millikelvin TEM are described. In addition, we provide a brief introduction to superconducting circuitry in the Appendix for the interested reader who is not familiar with the subject.

# 1. Introduction

In this paper, we argue that superconducting qubits are objects that would be both interesting *and* useful in TEM. On the other hand, superconducting qubits, which are one of the most promising physical systems for quantum information processing, demand a millikelvin environment. Thus, we contend that millikelvin TEM will profitably connect, in a nontrivial manner, research fields of quantum information technology [1], high-resolution biological electron microscopy [2], and quantum electron optics [3,4]. Below, we briefly review each of these subjects. Before proceeding, we direct the reader's attention to other recent proposals that emphasize the use of cryogenic TEM to investigate quantum materials [5-7].

Quantum information technologies (QITs) such as quantum cryptography [1], quantum computing [1] and quantum sensing [8] have attracted much attention in recent years. These technologies are characterized by being quantum mechanical at the *systems level*, rather than at the materials level. Theoretical insights, such as the existence of various quantum algorithms and error correction methods among others, stand by itself regardless of experimental development. On the other hand, progress in hardware technology, such as superconducting circuits [9], trapped ions [10] and light optics [11] among others, is also impressive. Possible applications of QIT in cryptography, materials simulation, machine learning, nanometer-scale sensing using the NV center etc. are widely believed to be important. Our contention here is that electron microscopy, although it currently is not attracting as much attention, could be one of the most important QIT applications.

We have already witnessed much progress in high resolution biological electron microscopy in recent years. Progress in cryoelectron microscopy (cryoEM) and single particle analysis (SPA) has enabled routine determinations of the structures of biological molecules [12-14]. However, we should note that the resolution of *raw* data still is highly limited by radiation damage to the specimens: One obtains a high-resolution structure only after averaging over many identical molecules. In other words, radiation damage prevents us from using a sufficient number of electrons for imaging, resulting in a large amount of shot noise that originates from the particle nature of the electron. On the other hand, as we will have determined all the relevant molecular structures at some point in the future, one may expect that the next focus of biological imaging should then be determination of locations and states of known molecules in the cellular environment. An important requirement here is that most molecules, as opposed to e.g. few fluorescently-labeled molecules here and there, should be comprehensively observable. Cryoelectron tomography (cryoET) has obvious relevance for such a task [15]. However, since we basically have only single specimens in cryoET (setting aside the method of subtomogram averaging [15] that still requires many identical molecules), we cannot perform averaging to improve signal-to-noise ratio (SNR) and the resolution remains to be $\approx 3-5$ nm [16]. This value falls just short of about $\approx 2-3$ nm that is needed for identification of protein molecules in the tomogram [17]. QIT involving free-space electrons could close this gap, which could be significant. Referring to cryoET, it has been mentioned that "If these methods develop to the point at which we can map macromolecules into 3D cell structure, a new era in cell biology is bound to emerge" [17].

This brings us to a discussion of quantum electron optics [3,4], where free-space electrons act as carriers of quantum information. Here, one may identify two distinct threads of research. First, much attention has recently been paid to experiments that involve manipulations of the quantum state of flying electrons through interactions with photons [18,19]. Such experiments have been made possible by increasing interaction between electrons and photons by e.g. an intense laser field [20] or by using artificial structures [21]. A point of the latter method is to engineer the dispersion curve of the photon to meet the condition of energy-momentum conservation upon electron-photon interaction [22]. Absorption or emission of photons by the electron results in a change of the electron energy. While in the present paper we are not interested in energy-changing processes, we will discuss interaction between an electron and *microwave* photons in later sections. In fact, microwave photons have been controlled with exquisite precision using superconducting electronics [23,24], which requires a millikelvin environment. Second, the use of quantum electron optics to improve imaging of beam-sensitive specimens, in particular biological specimens, has been discussed for some time now and experimental efforts began to be reported recently. For example, the use of interaction-free measurement in the context of electron optics has been proposed more than a decade ago [25] and recently its simple form has been experimentally demonstrated [26], using microfabricated gratings. Although interaction-free measurement is an obviously attractive approach to observe beam-sensitive specimens, it should also be noted that certain limitations with semi-transparent specimens are known [27,28]. Another idea is to gain certain advantage by manipulation of electron wave fronts by e.g. microfabricated diffraction gratings [29]. Experimental demonstrations begin to emerge in this area [30], while new technologies for electron wave manipulation emerge [31]. Yet another idea, *quantum electron microscopy* (QEM), is to go beyond the standard quantum limit to approach the Heisenberg limit [32]. This may be done either with a repeated use of single electrons [33,34], which is referred to as *multi-pass TEM*, or with the use of entanglement between an electron and superconducting qubits [35,36], which will be referred to as *qubit-assisted TEM*. The latter method, while not requiring unusual electron optics, does demand millikelvin temperature environment and it is this last possibility that we are going to study in the remaining sections of this paper.

## 2. An overview on instrumental configurations

The reason for building a millikelvin TEM, that we discuss in the present work, is the existence of superconducting qubits. See Appendix A for a brief introduction to superconducting qubits. Briefly, superconducting qubits are fabricated typically on a silicon substrate with technologies similar to ones found in semiconductor industry. The Josephson junction, the key element, is typically made of aluminum with the superconducting transition temperature $T_c \approx 1$ K. However, merely having the superconducting state is not good enough: The temperature $T$ must satisfy $k_B T \ll \Delta E$, where $\Delta E$ is the energy difference between relevant states of the superconducting qubits. The needed temperature turns out to be in the millikelvin region, which can be generated by the dilution refrigerator, or adiabatic demagnetization of certain materials.

An advantage of the superconducting qubit, when compared to other qubits, is that it admits various designs: The two states of a qubit are associated with distinct amounts of electrostatic charge in the case of the charge qubit, or distinct amounts of magnetic flux for the flux qubit, for example. Being a qubit, one can have quantum superpositions of these distinct charge/magnetic flux states: Analogous to the electron wavefunction with two wave packets at two distinct locations simultaneously, a superconducting qubit can have two distinct charge/magnetic state simultaneously. This is a remarkable property of the superconducting quantum circuit that is unheard of in classical electronics.

We propose to let the probe electrons fly by a superconducting qubit fabricated on a planer substrate in a TEM. The surface of the substrate is parallel to the optical axis of the TEM. We expect the distance between the electron and the qubit to be of the order of 1 μm. The trajectory of the electron is influenced by the Lorentz force, which comes from a quantum mechanically superposed electromagnetic field generated by the qubit. As a result, the electron and the qubit get quantum mechanically entangled. Mere observation of such an effect would be unprecedented. Moreover, applications of such phenomena are already in the pipeline, as we discuss in Sec. 4.

To observe entanglement of the probe electron and the superconducting qubit, we place the qubit as a specimen in a TEM. However, this will not be straightforward because this particular specimen must be cooled down to millikelvin temperatures. Expected engineering issues associated with combining a millikelvin environment and the TEM is discussed in Sec. 5. Once the electron and the qubit are entangled, the experimenter would be able to observe, e.g. the violation of the Bell inequality or related inequalities, by measuring both the electron and the qubit states. Ideally, pulsed single electron source would be needed for straightforward interpretation of the experimental results.

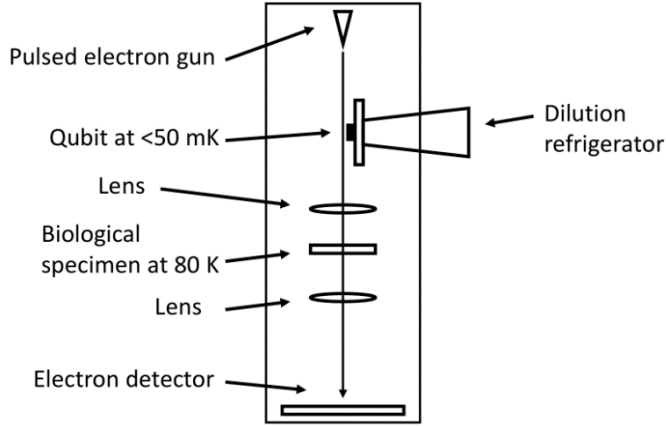

**Fig. 1.** A schematic design of a qubit-assisted TEM. The qubit is attached to the lowest temperature stage of a dilution refrigerator, which is shown schematically. Electrical wirings connecting the qubit and room-temperature electronics are not shown. The qubit and the electron detector are placed on planes that are conjugate to the back focal plane of the objective lens. The upper lens in the figure schematically represents the entire illuminating system including the pre-field of the objective lens, while the lower lens in the figure represents the downstream lens system that includes the objective lens and the projector lens.

On the other hand, each application of electron-qubit interaction would employ a case-specific experimental configuration. Figure 1 shows an experimental configuration of qubit-assisted TEM that goes beyond the standard quantum limit, or equivalently the shot noise limit. The most prominent application area of this is perhaps biological cryoEM, where the specimen is easily damaged by the electron beam and approaching the Heisenberg limit makes great sense. As detailed in Sec. 4, the probe electron first goes through the quantum mechanically superposed electromagnetic field, then gets focused by the condenser lens (perhaps combined with the pre-field of the objective lens) to interact with the specimen, and finally detected after going through the objective lens and other associated lens. The qubit and the electron detector are on their respective planes and both of these planes are conjugate to the back focal plane of the objective lens. As with other QITs, the reader may feel puzzled that we obtain the measurement result in the qubit while the electron interacts with the qubit *before* it interacts with the specimen. However, this "paradox" is resolved by noting that, depending on the measured electron state, we manipulate the qubit before measuring its state, as explained in Sec. 4.

We note that qubit-assisted TEM is not the only proposed way to approach the Heisenberg limit. Multi-pass TEM, in which the probe electron interacts with the specimen multiple times, has been proposed [33] (discussed also in Ref.[3] and, in passing, in Ref. [34]) in order to go beyond the standard quantum limit. Multi-pass TEM does not require the millikelvin temperature environment, but it comprises unconventional electron optics. See Sec. 4 for a preliminary comparison between qubit-assisted TEM and multi-pass TEM.

Another application of a superconducting qubit interacting with electrons, or indeed with any single-charged particles, is detection of such particles without affecting their trajectories so that they can continue flying [37]. This charged particle detector, while requiring millikelvin refrigeration, can be inserted at any place of a charged particle optics system.

## 3. Interaction between a flying electron and microwave photons

In order to have full entanglement, electron-qubit interaction needs to be such that the two qubit states, say $|0\rangle$ and $|1\rangle$, result in two distinguishable electron states. Appendix B reviews the case of flux qubit, where the electron trajectory is modified depending on the magnetic field trapped in the qubit, which in turn depends on the qubit state. Since the electron wave needs to be focused above the area of the qubit, the exit wave from the qubit has a natural angular spread due to diffraction. Hence the electron trajectory needs to be bent with an angle larger than the aforementioned angular spread, necessitating certain minimal amount of magnetic flux trapped in the qubit. This amount of magnetic flux turns out to be about the flux quantum $\phi_0 = \frac{h}{2e}$. This requires further design considerations because typical flux qubits tend to hold magnetic flux that is much less than $\phi_0$.

In this section, we consider another possibility of the use of microwave photon states in a resonator (Fig. 2). This idea is inspired by the development of circuit-QED systems in the field of superconducting quantum circuit [23,24,38], as well as the recent research activities in electron-photon interaction in the optical region in a TEM [18,19,21,22,31]. However, we focus on *microwave* photons instead of optical photons. One reason is that microwave photon quantum states have been controlled in circuit-QED systems exquisitely well [24], although it requires millikelvin temperatures. The interested reader can find comprehensive references to microwave-based qubits and their variants elsewhere [39]. We consider how the electromagnetic field associated with the photons could affect the trajectory of the electron flying through the field. Instead of discussing typical circuit-QED systems, where the transmon qubit is employed, we consider a much simpler system, which still keeps the core physics, for the purpose of order estimation. In particular, we model the microwave cavity of circuit-QED system with the simple LC lumped-circuit resonator. The temperature needs to satisfy $k_B T < \hbar \omega_c$, where $\omega_c$ is the resonant frequency. We will arrive at the conclusion that $\approx 100$ photons need to be stored in the resonator in order to sufficiently affect the electron trajectory. Incidentally, manipulation of $\approx 100$ photons in a superconducting cavity has already been demonstrated experimentally [24].

To consider photon-electron interaction in the simplest possible setting, we combine the LC resonator and the rf-SQUID. These are discussed in examples E1 in E2 of Appendix A, respectively. (We note that the resonator is usually modeled as a distributed microwave circuit in the circuit QED community, as opposed to the LC resonator comprising lumped elements.) The following analysis is very simple, elementary and semi-classical: We simply compute the electromagnetic field that would be produced at the instant when the LC resonator has its all energy in the magnetic or electrostatic field, and see if the field sufficiently deflects the electron flying by. We simply ignore the temporal dynamics of the electromagnetic field. We can store many photons in the resonator controllably and moreover we can make a quantum superposition of two different states, each containing many photons but oscillating with an opposite phase angle, i.e. with electric/magnetic fields pointing to opposite directions [23]. Superposition of many photons and no photons may also be generated. We take advantage of such a superposed quantum state of microwave photons to enable entanglement with flying electrons.

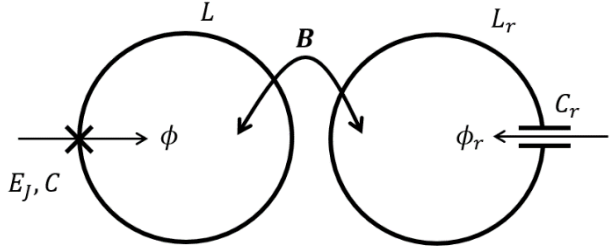

**Fig. 2.** A coupled LC resonator and rf-SQUID. This mimics the well-studied circuit QED system.

The Hamiltonian, which can be obtained by following the rules listed in Appendix A, has the form

$$H = \left\{\frac{q_r^2}{2C_r} + \frac{\phi_r^2}{2\tilde{L}_r}\right\} + \left\{\frac{q^2}{2C} + \frac{\phi^2}{2\tilde{L}} - E_J \cos\frac{2\pi\phi}{\phi_0}\right\} + \frac{\phi_r \phi}{L_c},$$

where the subscript $r$ denotes the resonator. The first group describes the resonator, the second represents the rf-SQUID, and the final term gives magnetic coupling between the two. Let $\omega_r$ be the resonant frequency $\frac{1}{\sqrt{\tilde{L}_r C_r}}$. Following the standard procedure, the resonator part of the Hamiltonian is expressed as $\hbar\omega_r a^\dagger a$ using the raising/lowering operators $a, a^\dagger$, neglecting the zero-point energy. These ladder operators obey the commutation relation $[a, a^\dagger] = 1$, The operator $\phi_r$ is expressed in terms of these as

$$\phi_r = \sqrt{\frac{\tilde{L}\hbar\omega_r}{2}}(a + a^\dagger).$$

Next, consider the rf-SQUID part. Because of the nonlinearity, one can limit the relevant Hilbert space to the one spanned only by the ground state $|0\rangle$ and the first excited state $|1\rangle$, with the energy difference $\hbar\omega$ between them. In short, we have a qubit. The raising/lowering operators here have the form

$$\sigma = \frac{\sigma_x - i\sigma_y}{2} = \begin{pmatrix} 0 & 0 \\ 1 & 0 \end{pmatrix}, \qquad \sigma^\dagger = \frac{\sigma_x + i\sigma_y}{2} = \begin{pmatrix} 0 & 1 \\ 0 & 0 \end{pmatrix},$$

which obeys the fermionic anticommutation relation $\{\sigma, \sigma^\dagger\} = 1$. The reader may verify that the energy of the qubit is expressed as $\hbar\omega\sigma^\dagger\sigma$. Note that the matrix forms of these operators equal that of $a, a^\dagger$ cut down to a 2-dimensional Hilbert space. Hence it is natural to assume that $\phi \propto \sigma + \sigma^\dagger$, in analogy with the operator $\phi_r$. (One may further convince oneself by noting that $\sigma + \sigma^\dagger = \begin{pmatrix} 0 & 1 \\ 1 & 0 \end{pmatrix}$, and also recalling that $\langle 0|\phi|0\rangle = \langle 1|\phi|1\rangle = 0$ because the ground state and the first excited state wavefunctions $\Psi(\phi)$ are even and odd respectively. Meanwhile, $\langle 1|\phi|0\rangle$ is nonzero, which we take to be real so that it equals $\langle 0|\phi|1\rangle$.) Thus, the Hamiltonian has the form

$$H = \hbar\omega_r a^\dagger a + \hbar\omega\sigma^\dagger\sigma + \lambda\Delta(a + a^\dagger)(\sigma + \sigma^\dagger),$$

where we defined $\Delta = \hbar(\omega - \omega_r)$ for later convenience and $\lambda$ is a small dimensionless constant. Treatment of this Hamiltonian is well-known [38]. Applying the rotation wave approximation (RWA), we further simplify the Hamiltonian as

$$H = \hbar\omega_r a^\dagger a + \hbar\omega\sigma^\dagger\sigma + \lambda\Delta I_+,$$

where $I_\pm = a^\dagger\sigma \pm a\sigma^\dagger$. This is the Jaynes-Cummings Hamiltonian (JCH) that is used in many areas of physics, including circuit QED [38].

We can diagonalize the JCH [40] using an easily verifiable relation

$$e^{-\lambda X} H e^{\lambda X} = H + \lambda[H, X] + \frac{\lambda^2}{2!}[[H, X], X] + \cdots.$$

Using a unitary operator $U = e^{\lambda I_-}$, we obtain, to second order in $\lambda$

$$H_{eff} = U^\dagger H U = (\hbar\omega_r - \lambda^2\Delta)a^\dagger a + \left[\hbar\omega + 2\lambda^2\Delta\left(a^\dagger a + \frac{1}{2}\right)\right]\sigma^\dagger\sigma$$

$$= \left[\hbar\omega_r + 2\lambda^2\Delta\left(\sigma^\dagger\sigma - \frac{1}{2}\right)\right]a^\dagger a + (\hbar\omega + \lambda^2\Delta)\sigma^\dagger\sigma.$$

Few remarks are in order. First, for small $\lambda$, the unitary operator $U$ is almost the identity operator. This means that the energy eigenstates are "weakly dressed" version of photons and the excited state of the qubit. Second, the expression for $H_{eff}$ in the first line reveals that the energy difference between the two qubit states depends on the photon number in the resonator. Likewise, the second line shows that the resonator frequency weakly depends on the qubit state. The latter has been used for measuring the qubit state. Furthermore, it has been demonstrated that one can populate the resonator with as many as $\approx 100$ photons only if the qubit state is $|0\rangle$ for example [24]. This means that we can generate a photonic state akin to $|\text{many photons}\rangle \otimes |0\rangle + |\text{no photon}\rangle \otimes |1\rangle$, if the initial qubit state is $(|0\rangle + |1\rangle)/\sqrt{2}$. (One can indeed disentangle the

qubit from the resonant cavity state, if so wished, by taking advantage of the fact that the qubit energy depends on the number of photons [23].) The experimentally demonstrated photon number $\approx 100$ is an interesting number because it is close to the inverse of the fine structure constant $\alpha^{-1} \approx 137$. The reason why $\alpha^{-1}$ is relevant comes from the way the resonator and the flying electron interact, to which we now turn.

Consider $n$ photons in an LC resonator. We consider the problem in a semi-classical manner. As noted before, the following calculation is for the purpose of rough estimation only and many aspects, such as temporal dynamics of the electromagnetic field associated with photons, will be ignored. The amount of energy stored is $n\hbar\omega_r = \frac{n\hbar}{\sqrt{\tilde{L}_r C_r}}$. This energy goes back and forth between the inductor and the capacitor. It appears that magnetic deflection of the electron trajectory is preferable because of the lack of energy exchange, but the problem is subtler than it appears, as will be described shortly.

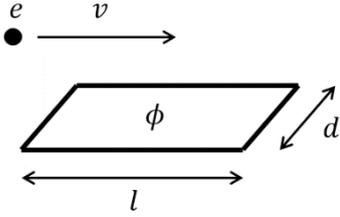

**Fig. 3.** A free-space electron flying by a superconducting circuit holding a magnetic flux $\phi$.

Consider magnetic deflection first. Let the shape of the inductor be the rectangular shape shown in Fig. 3. The trajectory of the flying electron is parallel to the long edge with length $l$. Our goal is to change the quantum state of the electron to a distinguishable state, which means that we want to deflect the electron beam with an angle of at least by $\frac{\lambda}{d}$, which is the beam spread due to diffraction. We are interested in the number of photons that are required to achieve a deflection by the angle $\frac{\lambda}{d}$. The needed magnetic flux is considered in Appendix B and it turns out to be $\phi = 2\phi_0$. At the moment when all the energy is in the inductor, the magnetic flux $\phi$ satisfies $\frac{\phi^2}{2\tilde{L}_r} = n\hbar\omega_r$. Hence, we need

$$n = \frac{\phi^2}{2\hbar\omega_r \tilde{L}_r} = \frac{4\phi_0^2}{2\hbar\omega_r \tilde{L}_r} = \frac{\pi R_K}{Z_r},$$

where $Z_r = \sqrt{\frac{\tilde{L}_r}{C_r}}$ is the characteristic impedance of the resonator, which tends to have a similar value with the impedance of the vacuum $Z_0 = \sqrt{\frac{\mu_0}{\varepsilon_0}} \approx 377$ Ω. We also note the fact that

$$\frac{2R_K}{Z_0} = \frac{2\varepsilon_0 hc}{e^2} = \alpha^{-1} \approx 137$$

Therefore, in order to deflect the electron sufficiently using the magnetic field, we need $\approx 100$ photons.

Next, consider deflecting the electron using an electric field. Consider a parallel plate capacitor with the rectangular plate with dimension $l$ by $d$, while the spacing between the plate is also $d$. For rough estimation purposes, we pretend that all the electric field lines are parallel between the two plates. Let the charge on the capacitor be $q$, then we obtain the electric field $E = \frac{q}{\varepsilon_0 dl}$. The deflection angle is

$$\theta = \frac{\Delta p}{p} = \frac{F\Delta t}{p} = \frac{eE}{p} \cdot \frac{l}{v} = \frac{eq}{p\varepsilon_0 dv} = \frac{h}{pd} \cdot \frac{eq}{\varepsilon_0 hv} = \frac{\lambda}{d} \cdot \frac{eq}{\varepsilon_0 h\beta c} = \frac{\lambda}{d} \cdot \frac{Z_0}{R_K \beta} \cdot \frac{q}{e} = \frac{\lambda}{d} \cdot \frac{2\alpha}{\beta} \cdot \frac{q}{e},$$

where $\beta = \frac{v}{c}$, which is $\approx 0.8$ for 300 keV electrons. Hence, we need

$$n_e = \frac{\beta R_K}{Z_0} = \frac{\beta}{2\alpha} \approx 55$$

electrons on the capacitor plate to sufficiently deflect the 300 keV flying electron. Next, we consider how many microwave photons are needed for this. When all the energy is in the capacitor, the charge $q$ satisfies $\frac{q^2}{2C_r} = n\hbar\omega_r$. Hence, we need

$$n = \frac{q^2}{2\hbar\omega_r C_r} = \frac{\beta^2 R_K^2 e^2}{2\hbar\omega_r C_r Z_0^2} = \frac{\pi\beta^2 R_K Z_r}{Z_0^2} \approx \frac{\pi\beta^2 R_K}{Z_0},$$

where the last approximation is valid when $Z_r \approx Z_0$. Under this approximation, the electrostatic case ostensibly has advantage by a factor $\beta^2 \approx 0.6$, in terms of the number of needed photons, over the magnetic case.

Here is a potential problem: When the flying electron is deflected by an electric field, the field does work on the electron. Such work may leave its trace in general, and hence the which-way information. Which-way information will prevent any quantum interference effect that we would like to measure or utilize later. The amount of work is computed to be, noting that the lateral motion of the electron is totally non-relativistic,

$$W = \int F dx = \int Fv dt = \int F \int a\, dt dt = \frac{F^2}{2m}\Delta t^2 = \frac{(eE)^2}{2m} \cdot \frac{l^2}{v^2} = \frac{1}{2\beta^2} \cdot \frac{(eEl)^2}{mc^2}$$

$$= \frac{1}{2\beta^2} \cdot \frac{\left(\frac{eq}{\varepsilon_0 d}\right)^2}{mc^2} = \frac{1}{2\beta^2} \cdot \frac{\left(\frac{e^2}{\varepsilon_0 d}\right)^2 n^2}{mc^2} \approx \frac{1}{2\beta^2} \cdot \frac{\left(\frac{e^2}{\varepsilon_0 d}\right)^2 \left(\frac{\pi\beta^2 R_K}{Z_0}\right)^2}{mc^2} \approx \frac{1}{2\beta^2} \cdot \frac{\left(\frac{e^2}{\varepsilon_0 d}\right)^2 \left(\frac{\pi\beta^2}{2\alpha}\right)^2}{mc^2}$$

$$= \frac{\pi^2 \beta^2}{8\alpha^2} \cdot \frac{\left(\frac{e^2}{\varepsilon_0 d}\right)^2}{mc^2}.$$

The first factor is rather large $\approx 10^4$, but we obtain $\frac{e^2}{\varepsilon_0 d} \approx 10$ meV for a realistic micro-fabricated resonator $d \approx 1$ μm. Since $mc^2 = 511$ keV, we obtain $W \approx$ μeV. On the other hand, the photon energy $\hbar\omega_r$ also is in the range of μeV if we use the microwave in the GHz range. Therefore, we might lose a photon to deflect the electron. This implies that we cannot use states with a definite photon number (i.e. the Fock states). Note that the experiment reported in Ref. [24] uses the coherent state, where the photon number does have uncertainty. As a final note, we remark that the which-way information will not necessarily be imprinted on the flying electron either. In order to have meaningful interaction with the oscillating electromagnetic field in the resonator, the electron beam must be pulsed, where the temporal width $\tau$ should be much shorter than the oscillating period $2\pi\sqrt{\tilde{L}_r C_r}$ of the resonator. The quantum uncertainty of the energy of such a pulsed electron beam, which is a fundamental uncertainty unlike the usual beam energy spread of an electron source, is

$$\Delta E = \frac{h}{\tau} \gg \frac{\hbar}{\sqrt{L_r C_r}} = \hbar\omega_r.$$

Hence the which-way information is deeply buried in uncertainty $\Delta E$ if $\hbar\omega_r \approx W$.

## 4. Applications of superconducting quantum circuits in TEM

Having seen ways to entangle a superconducting qubit and the flying electron, we proceed to briefly review and discuss possible applications of such phenomena. The first application A1 is qubit-assisted TEM, while the second application A2 is quantum nondemolition measurement of the lateral electron position.

**A1**) Qubit-assisted TEM [35,36] (Fig. 1): As mentioned in Sec. 1, QEM could contribute significantly to low-dose imaging of beam-sensitive biological specimens. Frozen biological specimens used in cryoEM are known to be weak-phase objects. Suppose two different regions A, B of the specimen induce phase shifts $\theta_A, \theta_B$ respectively to the probe electron wave. Consider the problem of detecting the phase difference $\delta = \theta_B - \theta_A$. For simplicity, we consider detection of the *existence* of non-zero $\delta$ and we do not care about the sign of $\delta$. A measurement that is able to determine the sign of $\delta$ has been described elsewhere [35,36].

We review measurement steps in qubit-assisted TEM. At any given moment, we have only two systems at most, that are the flying electron and a superconducting qubit. Furthermore, we consider only two quantum states of the electron $|0\rangle, |1\rangle$. A quantum state of the combined system is written as $|01\rangle$ for instance, which means that the electron is in the state $|0\rangle$, while the qubit is in the state $|1\rangle$. Figure 4 visualizes the entire state of the combined system of the electron and the qubit, for several measurement steps. The physical realization of the qubit is not important here. The qubit may be the flux qubit, one based on circuit QED, or something entirely different. We will often omit the unimportant overall factor of a quantum state, such as $\frac{e^{i\theta}}{\sqrt{2}}$, in the following discussion.

The followings are the measurement steps. First, the qubit is set to the state $|0\rangle + |1\rangle$. Second, we set the initial state of the probe electron from the electron source to $|0\rangle$. Figure 4 (a) shows the state of the combined system at this point. The system is set up in such a way that the electron beam gets deflected, via a magnetic field etc., to another state $|1\rangle$ if and only if the qubit is in the state $|1\rangle$. This would result in an entangled state $|00\rangle + |11\rangle$ (Fig. 4 (b)). Next, assume that the electron states $|0\rangle, |1\rangle$ are such that the state $|0\rangle$ corresponds to an electron wave going through the specimen region A; and $|1\rangle$ likewise corresponds to the region B. Now, suppose that region B induces more phase shift than region A, by an amount $\delta$. This results in the overall state as

$$|00\rangle + e^{i\delta}|11\rangle,$$

which is shown in Fig. 4 (c). Next, we apply a unitary operation $U$ to the electron state (Our focus here is to see what is possible *in principle*. Hence, we set aside the issue of practical realization), which converts $c_0|0\rangle + c_1|1\rangle$ into $c_0'|0\rangle + c_1'|1\rangle$, where

$$\begin{pmatrix} c_0' \\ c_1' \end{pmatrix} = U \begin{pmatrix} c_0 \\ c_1 \end{pmatrix} = \frac{1}{\sqrt{2}} \begin{pmatrix} 1 & 1 \\ 1 & -1 \end{pmatrix} \begin{pmatrix} c_0 \\ c_1 \end{pmatrix}.$$

In particular, up to the unimportant overall factor, the state $|0\rangle$ is converted to $|0\rangle + |1\rangle$ while the state $|1\rangle$ is converted to $|0\rangle - |1\rangle$. Hence, as shown in Fig. 4 (d), the application of the operation $U$ results in

$$(|00\rangle + |10\rangle) + e^{i\delta}(|01\rangle - |11\rangle) = |0\rangle(|0\rangle + e^{i\delta}|1\rangle) + |1\rangle(|0\rangle - e^{i\delta}|1\rangle).$$

This is followed by a measurement of electron state with respect to the measurement basis $|0\rangle, |1\rangle$. Suppose that the measurement outcome was $|0\rangle$. Then the qubit would be left in the state $|0\rangle + e^{i\delta}|1\rangle$. On the other hand, if the measurement out come turns out to be $|1\rangle$ (the probability for the two outcomes are both 50%), we obtain the qubit state $|0\rangle - e^{i\delta}|1\rangle$ (Fig. 4 (e)). Now, a well-known standard operation on a single qubit is the "phase gate" operation, defined as

$$|0\rangle \Longrightarrow |0\rangle, \qquad |1\rangle \Longrightarrow -|1\rangle.$$

(The operation is of course linear.) Applying this operation when the measurement outcome is $|1\rangle$, we can "correct" the qubit state to $|0\rangle + e^{i\delta}|1\rangle$ without knowing the value of $\delta$.

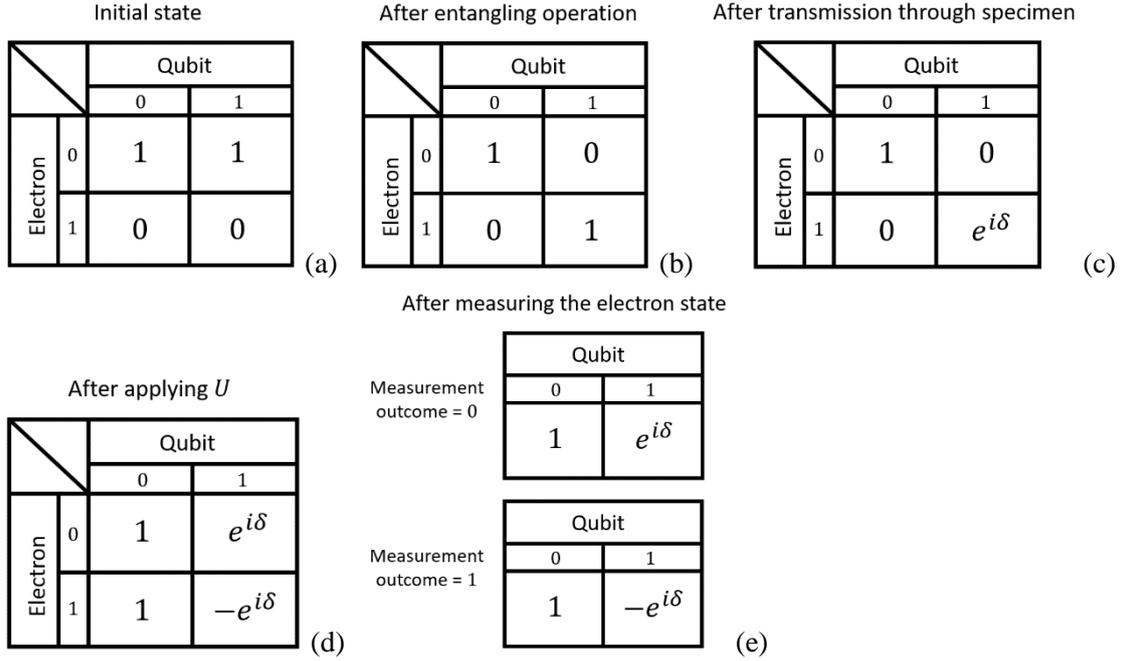

**Fig. 4.** The quantum state of the combined system of the electron and the qubit. Each of the electron and the qubit has two states here and therefore the combined system has four states. Hence, we show four amplitudes in each diagram. (The overall phase is arbitrarily fixed and the amplitudes in the diagrams are not normalized; and hence they are only proportional to the probability amplitudes.) (a) The initial state of the unentangled system comprising the electron and the qubit. (b) The state after the entangling operation. (c) The state after passing of the electron through the specimen in the TEM, receiving a phase shift. (d) The state after the unitary operation $U$ applied to the electron. (e) The state of the qubit after measuring the electron state, which depends on the measurement outcome.

We repeat the process with a new electron in the initial state $|0\rangle$. After entangling operation between the new electron and the qubit, the state of the combined system is $|00\rangle + e^{i\delta}|11\rangle$. Then this electron goes through the specimen, which results in the state $|00\rangle + e^{2i\delta}|11\rangle$. After an application of the operator $U$, we obtain the state

$$|0\rangle(|0\rangle + e^{2i\delta}|1\rangle) + |1\rangle(|0\rangle - e^{2i\delta}|1\rangle).$$

Notice the 2 on the exponent. Repeating this process $k$ times, using $k$ flying electrons, we eventually obtain the state

$$|0\rangle + e^{ik\delta}|1\rangle. \qquad (1)$$

(The astute reader may notice that, in fact, the above "correction" step needs to be performed only once in the end. In a sense, even no correction step at all would suffice, if we consider the entire process including interpretation of the measurement outcome of the qubit.)

Finally, we measure the qubit state. Since the overall phase factor does not matter in quantum mechanics, the state in Eq. (1) is equivalent to $e^{-\frac{ik\delta}{2}}|0\rangle + e^{\frac{ik\delta}{2}}|1\rangle$. Hence, application of the unitary operation $U$ results in

$$\cos\frac{k\delta}{2}|0\rangle - i\sin\frac{k\delta}{2}|1\rangle.$$

We measure the qubit state using the basis $|0\rangle, |1\rangle$. The measurement outcome $|1\rangle$ should signal the existence of the phase shift difference $\delta$ between the specimen regions A and B, which we wish to detect. The probability of detection is

$$p = \left|\sin\frac{k\delta}{2}\right|^2 \approx \frac{k^2\delta^2}{4}. \tag{2}$$

The approximation is valid if $k\delta \ll 1$, which we assume henceforth for simplicity.

To see the quantum advantage, compare this result with a classical measurement process using $k$ electrons separately. In classical measurement, each electron may be prepared in the state $|0\rangle + |1\rangle$, which becomes $|0\rangle + e^{i\delta}|1\rangle$ after transmission through the specimen. Hence for each measurement, the probability to detect the existence of non-zero phase difference $\delta$ is Eq. (2) with $k = 1$, that is $\frac{\delta^2}{4}$. Repeating the measurement $k$ times, probability of *not* detecting the existence of phase difference $\delta$ is

$$\left(1 - \frac{\delta^2}{4}\right)^k \approx 1 - \frac{k\delta^2}{4},$$

which means that the detection probability of phase difference is $\frac{k\delta^2}{4}$. This is worse than the quantum case in Eq. (2) by a factor $k$.

We remark on an alternative approach to realize QEM, i.e. multi-pass TEM [33,34]. Multi-pass TEM uses a single electron repeatedly, using an electron optics that refocus the same electron wave multiple times on the specimen. More specifically, we place two electron mirrors, which can quickly be removed when necessary, on both sides of the specimen. This would result in an accumulation of phase shift on the electron wave itself, allowing for the same quantum advantage with an arguably simpler setup. Since QEM is an active area of research and it is perhaps too early to tell which of the two, i.e. qubit-assisted TEM or multi-pass TEM, would be better for what purposes. Nonetheless, here we attempt to identify several pros and cons of the two methods. First, from the engineering perspective, it is highly challenging to realize both these methods. For instance, multi-pass TEM would need electron mirrors that can controllably be opened or shut on perhaps ns time scale *and* that desirably works at a potential $\approx 300$ keV or so. The reason is that higher acceleration energy is better in cryoEM, especially for relatively thick specimens such as a slice of frozen cell. Moreover, such specimens are expected to be important in the future. This circumstance is compounded by the fact that the *effective* thickness of specimen is $k$ times larger in QEM. (Somewhat tangentially, one may wonder if incident electrons coming from both sides of the specimen is problematic. However, we deal with weak phase objects at fairly low resolution at least in biological cryoEM and we do not expect much trouble in this regard.) Repetitive use of an electron using magnetic electron optics is also conceivable, but that seems to entail a whole different set of challenges. On the other hand, qubit-assisted TEM needs to incorporate millikelvin refrigerator into a TEM. Although there is one precedent combining a TEM and a dilution refrigerator for a different purpose, namely energy-dispersive X-ray spectroscopy [41], this will be challenging because of the need to insulate the TEM from the mechanical oscillation caused by the refrigerator, for example. Another disadvantage of the qubit-assisted TEM is the necessity of a better electron detector. In the measurement procedure discussed above, failing to detect even one electron out of $k$ electrons would result in a failure of the entire measurement. Hence, the probability of missing an electron must be about $k$ times smaller in the case of qubit-assisted TEM compared to multi-pass TEM. Yet another challenge for qubit-assisted TEM is the necessity to incorporate a pulsed electron beam, either by way of a pulsed electron gun or a beam blanker, which is synchronized with the qubit operation at the GHz time scale, although it does not have to be guaranteed that each pulse contains an electron. On a more fundamental note, however, an advantage of qubit-assisted TEM would be its potential extendability to allow for "universal" operation involving multiple pixels [37], by transferring quantum information back and forth between the TEM and a quantum information processor. Although a TEM controlled by a quantum computer might seem futuristic at the moment, the present authors do expect that a useful multipixel quantum measurement procedure exists for biological imaging, offering a much-needed "killer-app" of quantum computing.

**A2**) Quantum nondemolition measurement of electrons [37]: This idea comes from a rather simple observation. First, define the symmetric and antisymmetric states respectively as

$$|s\rangle = \frac{|0\rangle + |1\rangle}{\sqrt{2}}, \qquad |a\rangle = \frac{|0\rangle - |1\rangle}{\sqrt{2}}. \tag{3}$$

Note that

$$|0\rangle = \frac{|s\rangle + |a\rangle}{\sqrt{2}}, \qquad |1\rangle = \frac{|s\rangle - |a\rangle}{\sqrt{2}}.$$

The interaction between a flying electron and a superconducting qubit described above is such that the electron state $|0\rangle$ becomes $|1\rangle$ if and only if the qubit state is $|1\rangle$, by deflecting the electron trajectory. By slightly extending this operation, it is possible to arrange our scheme so that the electron state $|0\rangle$ becomes $|1\rangle$ *and* the electron state $|1\rangle$ becomes $|0\rangle$, if the qubit state is $|1\rangle$, while nothing happens if the qubit state is $|0\rangle$. (To realize this, one could divide the electron beam into two spatially localized states $|s\rangle$ and $|a\rangle$, and let the qubit interact with the electron wave in the latter state $|a\rangle$ only, so that the state receives a $\pi$ phase shift if and only if

the qubit state is $|1\rangle$. In fact, this was the original proposal for interaction between a flying electron and a qubit [35].)

The above operation, that the electron state flips as $|0\rangle \Leftrightarrow |1\rangle$ if and only if the superconducting qubit state is $|1\rangle$, turns out to be the paradigmatic operation in quantum information processing, i.e. the controlled-not operation (CNOT). In the present case, the superconducting qubit is the *control qubit*, while the electron is the *target qubit*. A remarkable, and well-known, fact in quantum information science is that the roles of control and target qubits are reversed, utterly unlike classical logic operations, if we see the same process in terms of the alternative basis defined in Eq. (3). That is, the superconducting qubit state flips as $|s\rangle \Leftrightarrow |a\rangle$ if and only if the electron state is $|a\rangle$. However, this means that one can detect the electron going through the state $|a\rangle$ in a quantum-nondemolition way. Unlike any existing single electron detector, this electron detector would allow one to use the very same detected electron in the down stream of electron optics. A simple idea, for example, is to place multiple such detectors in series, resulting in a lower error rate. This detector should work not only for electrons, but also for any single-charged particles such as ions and charged molecules. This is likely to have interesting applications, and an application to nano-fabrication has been put forward [37].

## 5. Engineering challenges in developing a millikelvin TEM

As we hinted in the above, building a qubit-assisted TEM presents an engineering challenge. In what follows, we attempt to list some of expected difficulties.

The first challenge is incorporation of a millikelvin environment in a TEM. First, we review standard precautions that we must follow in millikelvin measurements [42]. It is not sufficient to simply stick a millikelvin refrigerator to a TEM: For instance, simply sticking a refrigerator at the nominal operation temperature $\approx 4$ K may give the experimenter only $\approx 10$ K or so. It is necessary to understand physics of heat leakage to protect the millikelvin environment. The Stefan-Boltzmann law says that the thermal radiation heat is proportional to $T^4$, where $T$ is the temperature of the heat source. This highlights the importance of having metallic radiation shields, each anchored at an appropriate temperature stage of the refrigerator: A 60 K radiation shield, for example, would reduce the heat flux from the room temperature $\approx 300$ K by a factor $\left(\frac{300}{60}\right)^4 \approx 600$. Indeed, having multiple radiation shields is a necessity when setting up a millikelvin environment. At lower temperatures (roughly, lower than the LHe temperature), the $\propto T^4$ radiation becomes negligible and thermal conduction is the dominant mechanism. For instance, copper wires are unsuitable for conveying electrical signals because of the associated heat leak along them. On the other hand, good thermal contact between the device that one intends to cool and the refrigerator itself, i.e. the mixing chamber of a dilution refrigerator, becomes nontrivial in the millikelvin region. Furthermore, electrical signal lines are also transmission lines that are akin to wave guides, through which photons propagate. Room temperature thermal photons would be fully detrimental if they directly reach superconducting circuits. Investigators in the field of superconducting quantum circuit have long appreciated the importance of having an analog of multiple radiation shields along the signal lines. Such a "shield" usually takes the form of microwave attenuator at several temperature stages of the refrigerator. (In view of Kirchhoff's law of thermal radiation, an attenuator at 60 K will still emits 60 K microwave photons, which in turn need to be adsorbed by a 4 K attenuator, and so on.) These are all standard precautions for experiments at millikelvin temperatures.

There are new aspects when we handle flying electrons at millikelvin temperatures. First, we obviously need literal *holes* for the electron to enter and exit the millikelvin region. This is unlike optical experiments at cryogenic temperatures, where one could use a glass window *at low temperature* that is transparent for the optical photons of interest, but is opaque for the wavelength relevant to thermal radiation. In this case, the "cold window" blocks room temperature infrared photons while radiating only few infrared thermal photons. However, we believe that the need for literal holes is not an insurmountable problem. One reason is that the heat flux is not quantitatively large for the 300 K radiation. From the Stephan-Boltzmann constant, one finds that heat flux through a hole with an area $(10~\mu m)^2$ is $\approx 50$ nW, which is well within the ability of a dilution refrigerator of normal size. Another aspect is that the peak wavelength for 300 K radiation is about 10 μm. This means that a metallic mesh with the sieve size of e.g. 1 μm would effectively block the radiation in a way the Faraday cage works. On the other hand, it could be a challenge to align multiple holes of size $\approx (10~\mu m)^2$ on each layer of radiation shield on the electron optical axis. The problem is exacerbated by the fact that most materials shrink by $\approx$ 0.5% at cryogenic temperatures when cooled from the room temperature. This implies that we perhaps need an *in situ* method to align the holes. Finally, we add that stray infrared photons do harm the performance of superconducting qubit somewhat [43].

Other obvious challenges include vibrations produced by a dilution refrigerator, causing problems to a TEM; and stray magnetic fields, especially ac component of them, from the TEM that is problematic to superconducting qubits, which generally demand ultra-stable magnetic environment. It is perhaps fair to say that we seldomly use liquid helium as cryogen these days. Instead, a turnkey pulse-tube refrigerator is often used as a part of dilution refrigerator in the context of quantum computing. However, the pulse-tube refrigerator does generate mechanical vibration as well as acoustic sound, although it is quieter than another popular choice, i.e. the Gifford-McMahon cryocooler. Hence, it may take effort to integrate a dilution refrigerator and a TEM without causing serious problem of mechanical vibrations. We point out again that there is a precedent combining the dilution refrigerator and a TEM [41]. On the other hand, the dilution refrigerator is not the only way to get down to the millikelvin region. Recently, a millikelvin scanning tunneling microscope has been reported, which is cooled by adiabatic demagnetization refrigeration [44]. This system does involve a liquid helium bath. As for stray magnetic fields from the TEM, we may need to incorporate multiple layers of mu-metal passive shields at room temperature and superconducting shields at cryogenic temperatures. Active compensation schemes may be useful for reducing both mechanical noise and magnetic field noise.

## 6. Our current experimental effort towards a millikelvin electron microscope

Here we briefly report on our ongoing effort towards performing electron optics experiments at millikelvin temperatures. It should make sense to start small, because incorporating a millikelvin refrigerator into a TEM would be a large, complex, and expensive project. Hence, we plan to make an ultrasmall scanning TEM (STEM) that can be inserted into a commercial dilution refrigerator. The objective is to see *anything* at around a superconducting circuit at millikelvin temperature to gain experience that we need to proceed. The reason for developing a STEM, as opposed to a conventional TEM, is purely convenience: There is no simple 2-dimensional electron detector that works at millikelvin temperatures.

There is no precedent of building a TEM that works at millikelvin temperatures. In particular, we plan to place the entire electron optics at the millikelvin region and this comes with challenges of its own. Important issues include the need to find an electron gun and detector that work at such temperatures. It *is* conceivable to use a thermionic emission electron gun, although placing an object with temperature of few thousand kelvin in a millikelvin environment may give highly unusual impression. In fact, electron systems trapped over the surface of a superfluid helium have been investigated by using such a thermionic electron gun heated for a brief time period as an electron source [45]. Nonetheless, at this time we plan to use a cold field emission gun to produce low energy ($\approx 100$ eV) electrons. For an electron detector that works at millikelvin temperatures, we plan to simply use a Faraday cup, because at such a temperature almost all semiconducting devices would not be functional, because all carriers in the semiconductor would be frozen. Hence, our STEM design would be a rather "low-tech" one: It comprises a field emitter, a Faraday cup, an einzel lens, and an electrostatic deflector. Figure 5 (a)-(d) shows the overall view and some parts of a room temperature "mockup" version of the still-incomplete STEM that we are in the process of building and testing. We have not operated the instrument at present. As for the control and software part, we plan to employ the "NanoMi" open source TEM system to control our system [46]. In the near future, we plan to make a "miniature" version of the STEM shown in Fig. 5 (a), which we plan to first test at $2-4$ K using a pulse-tube refrigerator system (Fig. 5 (e)) before sending it down to millikelvin temperatures.

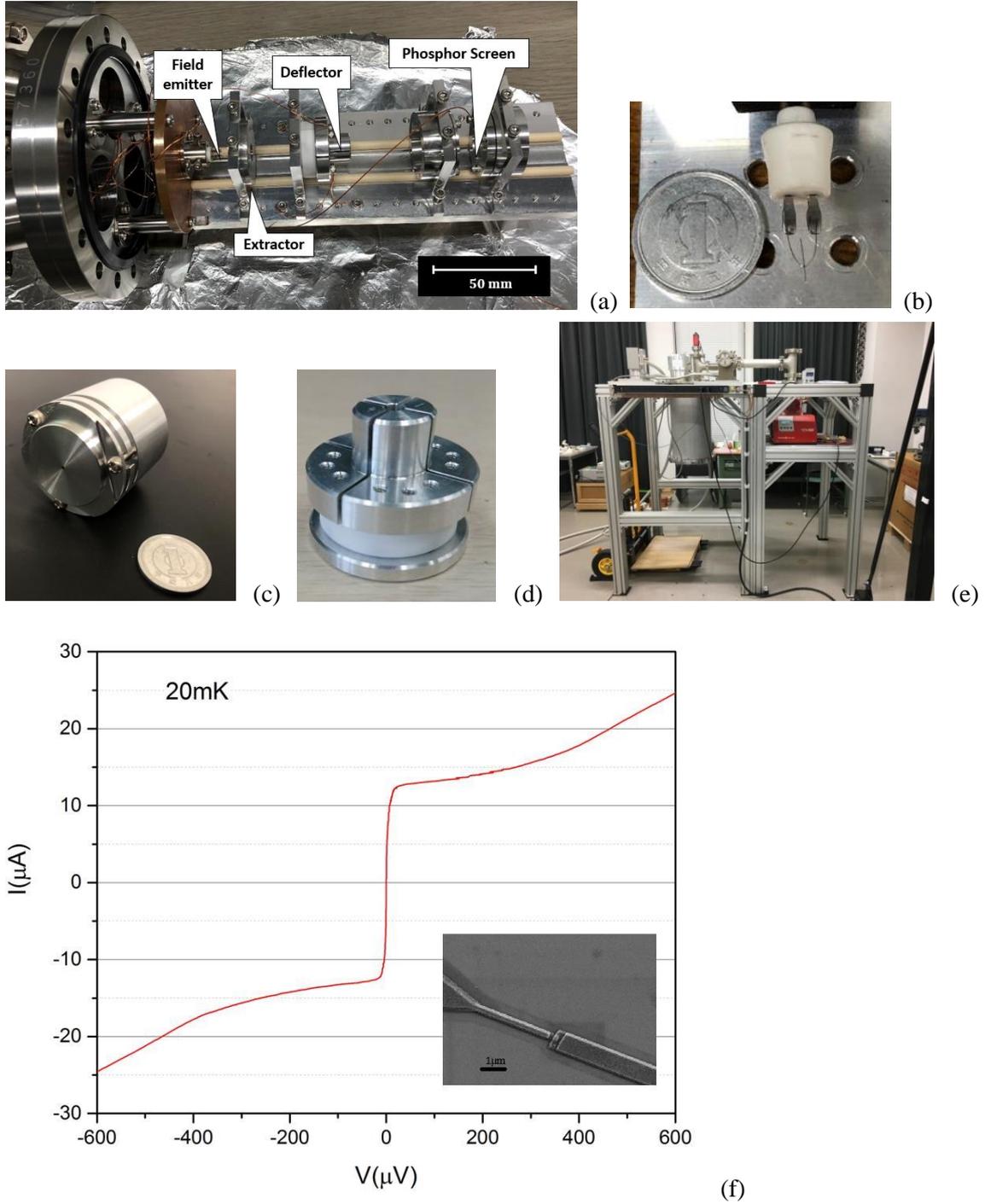

**Fig. 5.** (a) A room-temperature "mock-up" version of our STEM system under development. (b) A field emitter spot-welded on a heating loop. (c) An einzel lens. (d) An electrostatic beam deflector. (e) A pulse-tube cooling system capable of reaching down to 2 K without heat load. (f) I-V characteristics of an aluminum Josephson junction fabricated using the shadow evaporation method. Inset: An SEM image of the junction. Data taken from [48].

The configuration of our ultrasmall STEM is going to be as follows. The field emission electron gun is followed by an Einzel lens, which produces an electron beam that is focused on a specimen placed at $\approx 100$ mm from the lens. An electrostatic beam deflector placed in between the lens and the specimen scans the beam. A Faraday cup placed behind the specimen measures the transmitted electron current. Suppose, for example, that our specimen is a thin opaque film (essentially all films are opaque to $100$ eV electrons) with an array of regularly placed holes, as in the case of Quantifoil®. Then we should observe a periodically changing current at the Faraday cup as we scan the electron beam.

Electron optics that has anything to do with superconducting qubit should be able to detect a small beam deflection caused by the magnetic flux quantum $\phi_0 = h/2e \approx 2 \times 10^{-15}$ Wb. The deflection angle is (See Eq. (A3) of Appendix B)

$$\theta = \frac{\lambda}{2d},$$

where $\lambda \approx 1$ Å is the electron wave length in the present case of $100$ eV electrons, and $d \approx 1$ μm is the width of a "qubit", i.e. its length along an axis perpendicular to the optical axis. (We plan to use a mock-up qubit in initial experiments.) Hence, we obtain the deflection angle $\theta \approx 10^{-4}$ rad. Next, suppose that we place a mock-up qubit at near the Einzel lens. At the specimen, which is placed at $L \approx 100$ mm from the lens, the shift of the electron beam should be $L\theta \approx 10$ μm. This should be detectable as a slightly distorted STEM image of the array of regularly placed holes mentioned earlier, if the characteristic size of the array is sufficiently small. Obviously, the instrumental configuration just mentioned is only a first step towards developing a millikelvin TEM for investigating superconducting qubits. For example, we would need, among others, a pulsed electron beam and microwave injection lines to investigate the feasibility of microwave-based quantum electron optics. Looking further ahead, testing of a qubit-assisted TEM would likely require a simple room-temperature TEM such as NanoMi [46], to which a dedicated millikelvin refrigerator is installed.

We have started testing the feasibility of using a field emitter in the dilution refrigerator. We have found, at room temperature, that it is sometimes possible to actually produce an electron beam with a little over $100$ V extraction voltage, without bringing the field emitter and the extractor electrode too close. (Too close a distance might result in mechanical contact of these two parts upon cooling.) This is important because we intend to minimize the amount of modifications to the dilution refrigerator, which currently does not have signal lines that withstand high voltage. We have also found that the field emitter, once heated in a vacuum, can emit an electron beam again without heating, after bringing it outside the vacuum for a short time (on the order of hours). This aspect is also important because we intend to introduce a "cleaned" field emitter into the dilution refrigerator, whereafter we are not able to heat it for the cleaning purpose. The reason is that the signal wires of the refrigerator are not capable of withstanding $\approx 1$ A heating current, because minimization of heat leak requires the use of thin wires.

We have also started exploring fabrication of superconducting devices. An example of the Al-O-Al Josephson junctions which were fabricated by shadow evaporation technique [47] is shown in the inset of Fig. 5 (f). The device pattern is defined by electron beam lithography on a PMMA/MMA bilayer resist. The pattern is designed in such a way that a suspended opening is formed on the thin top layer of PMMA which defines the leads of the junction. The lower molecular weight MMA layer is innately over-developed to yield an exaggerated undercut which is needed for angled evaporation. In shadow evaporation technique the pattern is designed in such a way that when the evaporation source is sited off to sample normal the films overlap to form the junction over a small region which is typically about 100 nm in size. Evaporation of aluminum is done in a thermal evaporator which had a base pressure below $10^{-6}$ mbar and allowed displacement of the sample via rotation between two evaporation cycles. The first aluminum film is coated while the sample-to-source line made an angle of 15° with the sample normal. The chamber was then filled with high purity $O_2$ to a pressure of $1-10$ mbar to form a self-limiting oxide layer on the Al film which acted as the tunneling barrier. The second aluminum is coated with an incident angle of -15° after the sample is repositioned inside chamber. The graph in Fig. 5 (f) shows the I-V characteristics of a junction measured at 20 mK which displays Josephson effect in a superconducting weak link. Critical current is 15 μA which is high for a typical Al/AlOx/Al Josephson junction. The voltage gap is roughly 300 μV which agrees with Ambegaokar-Baratoff relation.

**7. Conclusions**

We described potential benefits of developing a millikelvin TEM. Although there could be many uses of such an instrument, we have highlighted the superconducting quantum circuits as possible specimens for such a TEM, because of the unusual electromagnetic fields that surround such devices. Such an electromagnetic field can be quantum superposition of presence *and* absence of a magnetic field, or that of a bunch of microwave photons. Observations of such electromagnetic fields will not only be scientifically unprecedented, but will also open avenues to new applications, which include imaging of radiation-sensitive biological specimens with unprecedented resolutions. We have also described expected engineering difficulties that we are likely to encounter when building a millikelvin TEM, and also described our ongoing attempt towards that goal. Since this direction of research requires knowledge of superconducting quantum circuit, with which electron microscopists are not necessarily familiar, we also provided Appendix A to serve as a brief and practical introduction to the subject area, for those who are interested. We also review interaction between a flux qubit and the flying electron in Appendix B.

**Acknowledgment**

The authors thank Mr. Shigeo Miura for fabricating a number of mechanical parts. We also thank Mr. Yuki Okuda, Mr. Yukihiro Takayama and numerous former undergraduate students, including Mr. So Kobayashi, for their early contributions to some of our experimental setups. We thank Dr. Marek Malac for useful discussions on the use of NanoMi open-source TEM, in particular the controller part, in our project. This research was supported in part by the JSPS "Kakenhi" Grant (Grant No. 19K05285).

**Appendix A. A brief introduction to superconducting quantum electronics**

Here we briefly review the bare minimum of superconducting quantum electronics from the user's perspective. We recall several facts first: A superconductor below the material-dependent transition temperature allows a lossless supercurrent in it, and also repels the magnetic field (the Meissner effect). To oversimplify, Cooper pairs of electrons, which are bosons with the charge $2e$, form Bose-Einstein condensate in a superconductor. (Pauli's principle forbids fermions to do the same.) Also recall that a Josephson junction (JJ), which arguably is the only active circuit element in superconducting electronics, comprises a thin insulating barrier between superconductors, which still allows a supercurrent to flow due to quantum tunneling (setting aside variants such as weak links). Other than these, we assume no background knowledge on the part of the reader, who we do assume to be familiar with analytical mechanics and quantum mechanics.

We introduce few rules for superconducting electronics without justification, from which everything we need is derived. Interested readers can look into relevant textbooks and reviews on topics such as the BCS theory, the Josephson effect, and superconducting qubits to see how our rules are related to the underlying physics [39,49]. Before proceeding, we remark that a "traditional" superconducting circuit such as the superconducting quantum interference device (SQUID), a magnetometer that include the JJs, is regarded as classical, despite that superconductivity as a phenomenon is very much quantum mechanical at the materials level. In contrast, a superconducting *quantum* circuit, which includes things such as ultrasmall JJs and qubits, behaves quantum mechanically at the *systems level*. We will see shortly that capacitances in the circuit plays a role analogous to masses in a mechanical system. This is the reason why submicron-scale JJs, with their small capacitance, is essential for quantum behavior, which is analogous to the fact that a mechanical system with light masses tends to behave quantum mechanically. Note that one of the leading technologies aiming at quantum computing is currently the superconducting quantum circuit technology [50].

We start with preliminaries. The magnetic flux quantum $\phi_0 = \frac{h}{2e} = 2.068 \times 10^{-15}$ Wb plays important roles in superconducting electronics. Another important constant is Klitzing's $R_K = \frac{h}{e^2} = 25.8$ kΩ. We deal only with supercurrent and will not consider lossy processes. We will use the lumped circuit approximation throughout this paper. We deal with superconducting wires, inductors, capacitors, JJs and nothing else. For analysis purposes, a large capacitor $C$ containing a large amount of charge $q$ may be regarded as a voltage source with the voltage $V = q/C$ irrespective of whether the actual voltage source is configured as such. Likewise, a large inductor $L$ trapping a large amount of magnetic flux $\phi$ may be regarded as a current source with the current $I = \phi/L$. A JJ is characterized by its characteristic Josephson energy $E_J$, which is typically measured in μeV and depends on the material and geometrical properties of each specific JJ. The critical current $i_c$, which is the maximal allowed supercurrent in a JJ, is given by $E_J = \frac{i_c \phi_0}{2\pi}$ as will be shown below. Each JJ inevitably has its junction capacitance $C$.

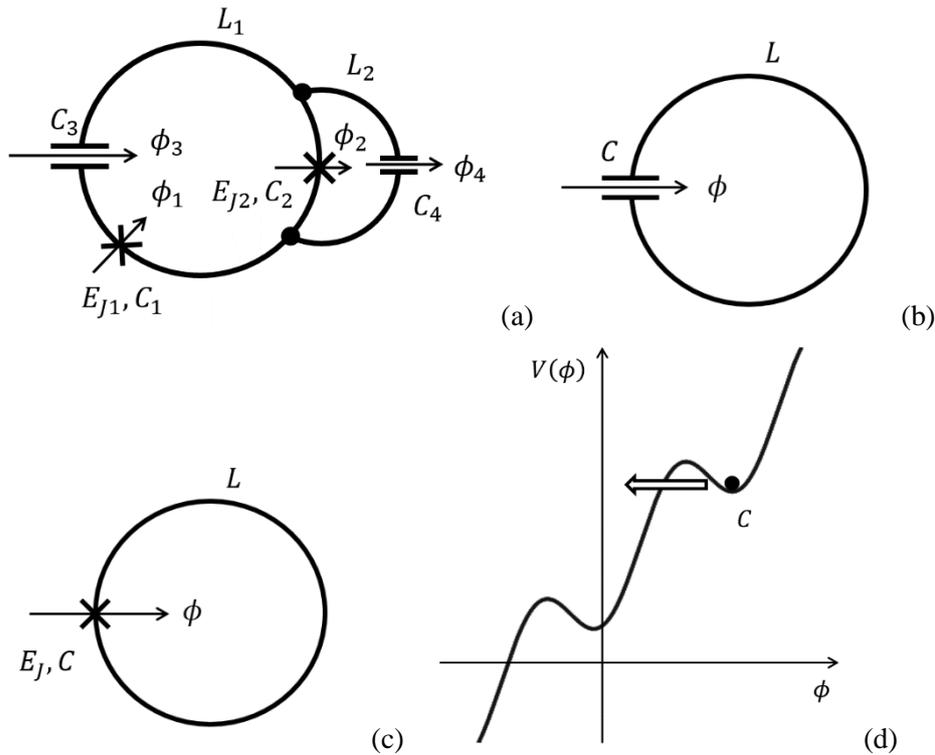

**Fig. A1.** (a) A general superconducting circuit. See the main text for the meanings of each symbol. (b) An LC resonator. (c) An rf-SQUID. (d) The "washboard" potential for a current-biased Josephson junction. The arrow illustrates a possible quantum tunneling process.

Figure A1 (a) shows an example of superconducting circuit. Crosses represent JJs and their associated capacitances. Inductors are shown simply as segments of curves, rather than by the usual symbol. Let a *loop* in a circuit be any closed curve comprising wires, inductors, JJs and capacitors. A capacitor may be considered as a JJ with infinitesimal $E_J$ (and hence infinitesimal supercurrent through it), with a specified junction capacitance. Because of the Meissner effect, the magnetic flux threading each "loop" is classically well-defined. When there is no JJs in a loop, (regarding a capacitor as JJ with zero $E_J$) the trapped magnetic flux cannot change because of the Meissner effect. However, in the presence of a JJ, where superconductivity is "weak", a magnetic flux can enter or exit the loop through the JJ. This prompts us to introduce the notion of the traversed magnetic flux (TMF) across a JJ, which represents the amount of magnetic flux that has crossed the JJ since a given reference point of time. More precisely, suggested by Faraday's law, the TMF through a JJ is *defined* as the time integral of the voltage across the JJ.

We are ready to describe our rules R1-R3 for superconducting circuits. We give the rules as a method to specify the classical Lagrangian of the circuit under consideration. Once a Lagrangian is obtained, one can analyze the system classically, or quantize it canonically as usual.

**R1**) Let each TMF $\phi_i$, which has traversed the $i$-th JJ, be a dynamical variable.

**R2**) By Faraday's law, the $i$-th JJ has a voltage $\dot\phi_i$ across the junction, resulting in the electrostatic energy $\frac{C_i}{2}\dot\phi_i^2$, where $C_i$ is the junction capacitance of the $i$-th JJ.

Remark: The term $\frac{C_i}{2}\dot\phi_i^2$ plays the role of kinetic energy in the Lagrangian because it involves a time derivative of the dynamical variable $\phi_i$. Since we regard pure capacitors as JJs with negligible $E_J$, a capacitor also provides a kinetic energy term.

**R3**) The potential energy part of the Lagrangian comprises two parts. First, the magnetic energy in a system of inductors have terms such as $\frac{\phi^2}{2L}$ and $\frac{1}{L_c}\phi_1\phi_2$, where the latter indicates coupling of inductors. Second, the energy stored in each JJ is given as $-E_{Ji}\cos\frac{2\pi\phi_i}{\phi_0}$, where $E_{Ji}$ is the Josephson energy of the $i$-th JJ.

Two remarks: First, $\theta_i = \frac{2\pi\phi_i}{\phi_0}$ is the (gauge-independent) superconducting phase difference across the JJ, which appears in many (if not all) textbooks in the field. Second, superconducting circuits tend to have only few degrees of freedom. Figuratively speaking, they are more akin to a few degrees-of-freedom small machines comprising rigid bodies, rather than a gas of electrons in metallic wires.

Having mentioned general rules, we proceed to examples E1-E4 to familiarize ourselves with superconducting circuits. We begin with a system that does not even include a JJ.

**E1)** LC Oscillator (Fig. A1 (b)): The TMF $\phi$ through the capacitor (which we regard as a JJ with $E_J = 0$) goes into the inductor, and only to the inductor. Thus, the TMF equals the magnetic flux $\phi$ trapped in the inductor. Hence the Lagrangian $\mathcal{L}$ (we use this symbol to distinguish it from the inductance) is given by

$$\mathcal{L} = \frac{C}{2}\dot\phi^2 - \frac{\phi^2}{2L}.$$

The conjugate variable to $\phi$ is the charge stored in the capacitor $q = \frac{\partial \mathcal{L}}{\partial \dot\phi} = C\dot\phi$. Hence, we obtain the Hamiltonian

$$H = \dot\phi q - \mathcal{L} = \frac{q^2}{2C} + \frac{\phi^2}{2L}.$$

The condition for canonical quantization is $[\phi, q] = i\hbar$, which leads us to the Schrodinger equation

$$i\hbar \frac{\partial \Psi(\phi)}{\partial t} = \left\{ -\frac{\hbar^2}{2C}\frac{\partial^2}{\partial \phi^2} + \frac{\phi^2}{2L} \right\} \Psi(\phi).$$

Unlike the textbook Schrodinger equation, the motion described by this equation is not in the real space $x$, but in an abstract space of magnetic flux $\phi$. Analogous to the fact that a wavefunction of an electron $\Psi(x)$ can be superposition of two wave packets located at distinct locations, it is possible to have an inductor trapping superposition of two different amounts of magnetic flux. It would be nice to "see" such a state with an electron microscope, but one must be careful because observations generally "collapse" the wavefunction $\Psi(\phi)$.

A few remarks follow. The above LC oscillator is the electronic analog of the mechanical harmonic oscillator, whose Lagrangian, Hamiltonian, and the canonical quantization condition are respectively $\mathcal{L} = \frac{m}{2}\dot x^2 - \frac{k}{2}x^2$, $H = \frac{p^2}{2m} + \frac{k}{2}x^2$, and $[x, p] = i\hbar$ expressed with the usual symbols. As remarked earlier, we see that the capacitance $C$ corresponds to the mass $m$ in the mechanical system. Also note that the energy quantum $\hbar\omega = \frac{\hbar}{\sqrt{LC}}$, where the frequency $\frac{1}{2\pi\sqrt{LC}}$ is typically measured in GHz, can be made larger than the thermal energy $k_B T$ by employing the $^3$He/$^4$He dilution refrigerator, which generates temperatures in the mK range. Furthermore, the gap energy $\Delta \approx 170$ μeV of aluminum, with which one can make high-quality ultrasmall JJs, is such that one can satisfy

$$k_B T < \hbar\omega < \Delta.$$

It is a rather fortunate circumstance that we can satisfy this formula, for the frequencies $\omega$ accessible with the well-developed microwave technology [51], that allows us to operate superconducting circuits in the quantum region without exceeding the energy scale $\Delta$ that governs physics of superconductivity, including that of JJs.

**E2**) rf-SQUID (Fig. A1 (c)): This is a type of SQUID magnetometer that was popular especially in the early days of superconducting electronics [52]. Today in QIT, the same system is sometimes used as a qubit, but still may be referred to by the old name of rf-SQUID. The only difference from the above example E1 is the non-zero $E_J$. Adding the potential term mentioned in the rule R3, we obtain the Hamiltonian

$$H = \frac{q^2}{2C} + \frac{\phi^2}{2L} - E_J \cos\frac{2\pi\phi}{\phi_0}. \tag{A1}$$

We see that the JJ introduces nonlinearity in the circuit. This enables much richer behaviors of superconducting circuit. For starters, the energy levels are no longer equally spaced. This enables us to use the system as a qubit because, for example, irradiation of a well-tuned microwave now allows for transition between only two states, that are the ground state and the first excited state of the system. The next two examples are variations of this system.

**E3**) Superconducting ring: A simple ring made of superconductor may be thought of as an rf-SQUID with a JJ with a very high critical current and hence a large $E_J$. This allows us to ignore the second term in Eq. (A1), resulting in a purely sinusoidal potential. The large $E_J$ then forces $\phi$ to be integer multiple of the flux quantum $\phi_0$, which is known as quantization of magnetic flux. It is the current in the ring that adjusts the value of $\phi$ to a quantized value.

**E4**) Current-biased JJ: This system is not directly relevant to our discussion of quantum electron optics applications, but we discuss this because of its simplicity and instructiveness. Consider the exact same system as above E2, but let the inductance $L$ be large. Furthermore, let the large inductor trap a large amount of magnetic flux $\Phi$ initially. Since the TMF through the only JJ is $\phi$, the magnetic flux in the inductor is $\Phi + \phi$. Since $\Phi \gg \phi$, the Hamiltonian is, neglecting an unimportant constant term

$$H = \frac{q^2}{2C} + \frac{(\Phi + \phi)^2}{2L} - E_J \cos\frac{2\pi\phi}{\phi_0} \approx \frac{q^2}{2C} + \frac{\Phi}{L}\phi - E_J \cos\frac{2\pi\phi}{\phi_0}.$$

However, a large inductor $L$ trapping a large magnetic flux $\Phi$ can be regarded as a current source with the current $I_b = \frac{\Phi}{L}$. Thus, the Hamiltonian of a current-biased JJ with the bias current $I_b$ is $H = \frac{q^2}{2C} + V(\phi)$, where the so-called "washboard" potential energy is given as (See Fig. A1 (d))

$$V(\phi) = I_b \phi - E_J \cos \frac{2\pi \phi}{\phi_0}.$$

Consider classical dynamics of this system. A fictitious particle with a "mass" $C$ moves in the washboard potential $V(\phi)$. Faraday's law states that $\dot{\phi}$ gives the voltage generated across the biased JJ. Imagine slowly increasing $I_b$ from the initial value of zero. Until certain point, the "particle" in the washboard potential does not start moving, meaning that we have zero voltage generated for a finite $I_b$. In other words, we have a supercurrent until the current reach a certain point, which turns out to be the critical current $I_c = \frac{2\pi E_J}{\phi_0}$. (Once the "particle" does start moving and a nonzero voltage appear, then dissipation mechanisms kick in, preventing an indefinite increase of the "velocity" of the particle. We do not go into the details here.)

Next, consider the quantized system. In the same procedure of increasing $I_b$ from zero, we see that the voltage should start manifesting itself when $I_b$ is still slightly below $I_c$. This is because the "particle" can quantum mechanically tunnel through a low potential barrier (Fig. A1 (d)). This is called macroscopic quantum tunneling (MQT) because we are talking about a collective motion of many electrons here [53]. Historically, MQT provided the first evidence that superconducting circuits behave quantum mechanically at the systems level [54,55].

**Appendix B. Review of interaction between a flying electron and a flux qubit**

Here we review consequences of the electron going through the magnetic field generated by the flux qubit (See Fig. A2 (a)). This results in bending of the trajectory of the electron due to the Lorentz force. This is yet another variation of example E2 of Appendix A, i.e. the rf-SQUID. We magnetically couple the rf-SQUID to a superconducting loop $L'$ with a persistent current $I'$, that traps a magnetic flux $\Phi$ forever as discussed in E3 in Appendix A. Our intention is to apply a bias magnetic field to the rf-SQUID, which results in half the magnetic flux quantum $\phi_0/2$ in the SQUID loop. The Hamiltonian of the system has the form

$$H = \frac{q^2}{2C} + \frac{\phi^2}{2\tilde{L}} + \frac{\Phi^2}{2\tilde{L}'} + \frac{\phi \Phi}{L_c} - E_J \cos \frac{2\pi \phi}{\phi_0}.$$

With a suitable adjustment of $\Phi$, neglecting an unimportant additive constant, we obtain

$$H = \frac{q^2}{2C} + \frac{(\phi - \phi_0/2)^2}{2\tilde{L}} - E_J \cos \frac{2\pi \phi}{\phi_0},$$

with an effective inductance $\tilde{L} \approx L$ for weak coupling between the two inductors. By shifting the variable to redefine $\phi$ as $\phi - \phi_0/2 \Rightarrow \phi$, the Hamiltonian may be expressed more simply as

$$H = \frac{q^2}{2C} + \frac{\phi^2}{2\tilde{L}} + E_J \cos \frac{2\pi \phi}{\phi_0}, \tag{A2}$$

This results in the potential curve shown in Fig. A2 (b), which has two minima at $\phi = \phi_A$ and $\phi_B$. In some sense, the rf-SQUID "has two choices" to make the trapped magnetic flux close to an integer multiple of the flux quantum $\phi_0$. One can use quantum states in these two minima as two qubit states. These states correspond to two different amounts of magnetic field in the inductor. This would apply two different amounts of Lorentz force to an electron flying nearby, bending the trajectory in different ways. Hence this interaction would generate entanglement between the qubit and a flying electron in an electron microscope. (We remark that these two states are not energy eigenstates. Their addition and subtraction are the energy eigenstates, with the energy difference related to the tunneling probability through the potential barrier at the center of Fig. A2 (b).)

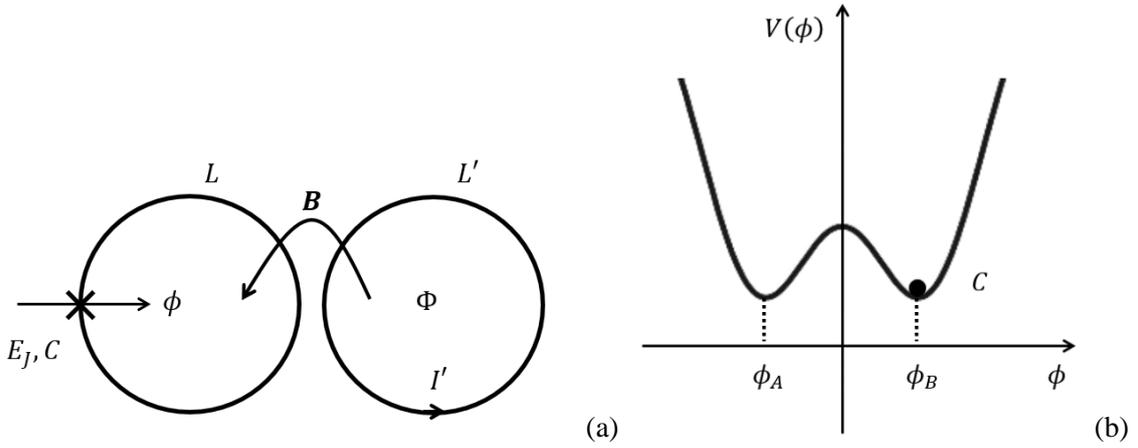

**Fig. A2.** (a) A flux qubit. It comprises an rf-SQUID and a superconducting persistent-current magnet that gives a bias magnetic field **B**. (b) Potential curve for the flux qubit. It has two minima, corresponding to two qubit states.

Consider the magnetic flux difference between the two qubit states $\Delta\phi = \phi_B - \phi_A$. One naturally tries to adjust the magnitudes of the second and third terms of Eq. (A2) properly, in order to make the height and width of the potential barrier between the two minima $\phi_A, \phi_B$ small enough to allow for qubit operations. However, this means that $\Delta\phi$ is smaller, and in fact typically much smaller, than $\phi_0$. However, below we see that $\Delta\phi \approx \phi_0$ is required to fully change the quantum state of the flying electron.

Figure 3 of the main text shows an electron flying near a rectangular coil containing a magnetic flux $\phi$. The dimension of the coil is $l$ times $d$ and hence the magnetic flux density is $B = \frac{\phi}{ld}$, resulting in the Lorentz force $F = evB = \frac{ev\phi}{ld}$, where $v$ is velocity of the flying electron. Newton mentioned that the momentum change of the electron is $\Delta p = F\Delta t$, where the time of interaction is essentially $\Delta t = \frac{l}{v}$. Putting things together, we obtain the deflection angle of the electron:

$$\theta = \frac{\Delta p}{p} = \frac{e\phi}{pd} = \frac{h}{2pd} \cdot \frac{\phi}{\phi_0} = \frac{\lambda}{2d} \cdot \frac{\phi}{\phi_0}, \tag{A3}$$

where $p$ is the momentum of the flying electron and $\lambda = \frac{h}{p}$ is the electron wavelength. On the other hand, in order to have a distinct quantum state of the electron, this deflection angle must be larger than the angular spread due to diffraction $\frac{\lambda}{d}$ that we already have. This gives us a condition

$$\phi > 2\phi_0.$$

Further considerations involving the Aharonov-Bohm effect shows that actually $\phi > \phi_0$ suffices to induce a distinct electron state [36]. In any event, we need a magnetic flux of the order of $\phi_0$ or more to properly affect the trajectory of the flying electron. The conceptually simplest way, which is not necessarily the simplest in practice, may be the use of fully entangled multiple flux qubits. There is also a theoretical proposal to realize such entanglement without the necessity of fully controllable multiple quantum bits [56]. However, the scheme appears to be somewhat complex.